# GAMMA SHAPE MIXTURES FOR HEAVY-TAILED DISTRIBUTIONS

By Sergio Venturini, Francesca Dominici and
Giovanni Parmigiani

*Università Luigi Bocconi, Johns Hopkins School of Public Health
and Johns Hopkins University*

An important question in health services research is the estimation of the proportion of medical expenditures that exceed a given threshold. Typically, medical expenditures present highly skewed, heavy tailed distributions, for which (a) simple variable transformations are insufficient to achieve a tractable low-dimensional parametric form and (b) nonparametric methods are not efficient in estimating exceedance probabilities for large thresholds. Motivated by this context, in this paper we propose a general Bayesian approach for the estimation of tail probabilities of heavy-tailed distributions, based on a mixture of gamma distributions in which the mixing occurs over the shape parameter. This family provides a flexible and novel approach for modeling heavy-tailed distributions, it is computationally efficient, and it only requires to specify a prior distribution for a single parameter. By carrying out simulation studies, we compare our approach with commonly used methods, such as the log-normal model and nonparametric alternatives. We found that the mixture-gamma model significantly improves predictive performance in estimating tail probabilities, compared to these alternatives. We also applied our method to the Medical Current Beneficiary Survey (MCBS), for which we estimate the probability of exceeding a given hospitalization cost for smoking attributable diseases. We have implemented the method in the open source `GSM` package, available from the Comprehensive `R` Archive Network.

**1. Introduction.** In health services research, there is extensive literature on models for predicting health costs or health services utilization. These prediction problems are usually complicated by the nature of the distributions being analyzed: high skewness, heaviness of the right tail and significant

---

fractions of zeros or token amounts are commonly encountered. At present, there is no agreement about the best methods to use [see Mullahy and Manning (2001), Kilian et al. (2002), Buntin and Zaslavsky (2004), Barber and Thompson (2004), Manning and Mullahy (2005), Powers et al. (2005), Dodd et al. (2004); for a recent survey, see Willan and Briggs (2006)].

An important and still open research question is how to best predict the proportion of (total or single-event related) medical expenditures that will exceed a given threshold [see, e.g., Briggs and Gray (2006), Conwell and Cohen (2005)]. For example, insurance companies and governmental health departments are often interested in predicting how many customers or citizens will ask for a reimbursement above a certain threshold. Similarly, financial institutions are often interested in estimating the probability of the potential loss that could take place in the next day, week or month. In all these situations the parameter of interest is a tail probability of a highly skewed distribution. Thus, it is important to develop methods that do not simply smooth the distribution of the data, but that are able to perform well from a predictive point of view.

The work presented in this article is motivated by an analysis of medical expenditures from the Medicare Beneficiaries Survey (MCBS). We are interested in modeling the distribution of medical costs paid by the Medicare program for treating smoking attributable diseases, specifically lung cancer (LC) and coronary heart disease (CHD). We need to estimate the probability that the hospitalization cost for a smoking attributable disease exceeds a certain value.

MCBS is a continuous, multipurpose survey of a U.S. nationally representative sample of Medicare beneficiaries (people aged 65 or older, some people under age 65 with disabilities and people with permanent kidney failure requiring dialysis or a kidney transplant). The central goal of MCBS is to determine expenditures and sources of payment for all services used by Medicare beneficiaries. The data set includes medical expenditures for LC or CHD as primary diagnosis for 26,834 hospitalizations of 9,782 individuals for the period 1999–2002. For our analyses, we extract medical expenditures on the first hospitalization for 7,615 individuals over the same period.

A typical assumption in health services research is that medical costs are log-normally distributed [Zhou, Gao and Hui (1997), Tu and Zhou (1999), Zhou et al. (2001), Briggs et al. (2005)]. In our case, as well as many others, this assumption is not appropriate, since the distribution of log-transformed expenditures is still far from being symmetric. For this reason, new methods have been recently proposed for estimating the cost mean difference between two groups [Johnson et al. (2003), Dominici et al. (2005), Dominici and Zeger (2005)]. However, few methods have been evaluated for modeling the entire distribution and for prediction.

The presence of few large observations is characteristic of skewed, heavy-tailed distributions. It is well known that these observations can influence the results of statistical analyses. The remedies proposed in the health research literature are either to transform the data [see Duan (1983), Mullahy (1998), Manning (1998), Mullahy and Manning (2001)] or to use robust methods [see Conigliani and Tancredi (2005), Cantoni and Ronchetti (1998)]. However, the combination of a transformation with a parametric form may still not be sufficiently flexible to accommodate large values, and also retains the undesirable property of borrowing information from the left tail to fit the right tail. On the other hand, robust methods may discount extreme outliers, even though those may be critical in helping decision makers evaluate the change of incurring in a very large cost. We sought a more reasonable trade-off by modeling the medical costs distribution using mixture, a direction that has not been explored in health services research as far as we know. Mixture models are parametric models which are flexible enough to represent a large spectrum of different phenomena. The mixture models literature is extensive [for general overviews, see Titterington, Smith and Makov (1985), Lindsay (1995), McLachlan and Peel (2000); for a comprehensive list of applications, see Titterington (1997)]. Particularly relevant for this paper is the well-developed parametric Bayesian literature on mixture distributions [see Diebolt and Robert (1994), Robert (1996), Roeder and Wasserman (1997), Marin, Mengersen and Robert (2005)].

Motivated by these considerations, we propose a new method for density estimation of very skewed distributions and for predicting the proportion of medical expenditures that exceed a given threshold. We model the distribution of medical expenditures by use of a mixture of gamma density functions with unknown weights. Using this model, we then estimate the tail probability $\mathbb{P}(Y > k)$, for different values of $k$. Each gamma distribution in the mixture is indexed by a component-specific shape parameter and a single unknown scale parameter $\theta$. This parameterization allows to create a convenient and flexible model characterized by a single parameter for all the gamma components, plus the ordinary set of mixture weights. Moreover, our parameterization overcomes the so-called "label-switching" identifiability problem that affects mixture model estimation by automatically providing an ordering of the mixture components. For a recent survey on identifiability problems in Bayesian mixture modeling, see Jasra, Holmes and Stephens (2005). The number of mixture components in our model is fixed. We provide practical advice on how to choose it as well as the hyperparameters of the prior on the scale parameter. In a simulation study closely mimicking the MCBS data we demonstrate that our method has a better predictive performance compared to standard approaches.

In Section 2 we introduce the gamma shape mixture model, the estimation approach, and provide guidance on how to choose prior hyperparameters. In Section 3 we illustrate the results of the simulation study and the data analysis. Section 4 contains a discussion and concluding remarks. The Appendix contains technical details about the Gibbs sampler used.

**2. The gamma shape mixture model.** In this section we introduce the gamma shape mixture (GSM) model. We begin by describing the likelihood and its main properties. In particular, we show that the gamma mixture model does not suffer from identifiability problems. We then present the prior structure and an approach for posterior calculations.

2.1. *Likelihood and prior structure.* Let $Y$ be a positive random variable, for example, nonzero medical expenditures. The GSM model is defined as

$$(2.1) \qquad f(y|\pi_1, \ldots, \pi_J, \theta) = \sum_{j=1}^{J} \pi_j f_j(y|\theta),$$

where $f_j(y|\theta) = \frac{\theta^j}{\Gamma(j)} y^{j-1} e^{-\theta y}$, the density function of a gamma $\mathcal{G}a(j, \theta)$ random variable. We assume that the number of components $J$ is known and fixed, whereas $\boldsymbol{\pi} = (\pi_1, \ldots, \pi_J)$ is an unknown vector of mixture weights. Discussion on how to choose $J$ will be provided later. In what follows, we denote (2.1) as $\mathcal{GSM}(\boldsymbol{\pi}, \theta | J)$.

The GSM model has two useful properties:

1. $\frac{1}{\theta}$ is a scale parameter [Lehmann and Casella (1998)] for the whole model, since

$$f(y|\pi_1, \ldots, \pi_J, \theta) = \theta \cdot f(\theta \cdot y|\pi_1, \ldots, \pi_J, 1).$$

2. Its moments are convex combinations of the moments of the $Y_j|\theta \sim f_j(y|\theta)$ mixture components, so that the $m$th moment is given by

$$\mathbb{E}[Y^m|\pi_1, \ldots, \pi_J, \theta] = \sum_{j=1}^{J} \pi_j \mathbb{E}[Y_j^m|\theta] = \sum_{j=1}^{J} \pi_j \frac{\prod_{\ell=1}^{m} (j + \ell - 1)}{\theta^m}.$$

A further issue related to mixture modeling is *label switching*, that is, invariance to permutations of the components' indexes [see Jasra, Holmes and Stephens (2005)]. A typical solution is to impose an *identifiability constraint*, usually an ordering of either the components means or the variances or the mixture weights [see Aitkin and Rubin (1985)]. A nice feature of the GSM model (2.1) is that it automatically imposes a constraint on both the means and the variances, since

$$\frac{1}{\theta} < \frac{2}{\theta} < \cdots < \frac{J-1}{\theta} < \frac{J}{\theta} \quad \text{and} \quad \frac{1}{\theta^2} < \frac{2}{\theta^2} < \cdots < \frac{J-1}{\theta^2} < \frac{J}{\theta^2}.$$

Therefore, the model is always identified and label switching is not a concern.

We assume that $\theta$ and $\boldsymbol{\pi}$ are independent a priori and we specify the following conjugate prior distributions:

$$\theta \sim \mathcal{G}a(\alpha, \beta),$$

$$\boldsymbol{\pi} = (\pi_1, \ldots, \pi_J) \sim \mathcal{D}_J\left(\frac{1}{J}, \ldots, \frac{1}{J}\right).$$

In practice, choosing the $\boldsymbol{\pi}$ prior hyperparameters equal to $1/J$ tends to produce posterior distributions where only a small subset of the $J$ mixture weights will have high prior probability to be selected at each iteration of the MCMC, as we will see in the application.

Given a sample $\mathbf{y} = (y_1, \ldots, y_n)$ of i.i.d. observations from (2.1), the likelihood is given by

$$(2.2) \qquad \mathbb{L}(\boldsymbol{\pi}, \theta | \mathbf{y}) = \prod_{i=1}^{n} \sum_{j=1}^{J} \pi_j f_j(y_i | \theta).$$

Unfortunately this expression is untreatable because it includes $J^n$ different terms [Marin, Mengersen and Robert (2005)]. To overcome this hurdle we use the so-called *missing data* representation of the mixture [Dempster, Laird and Rubin (1977), Tanner and Wong (1987), Diebolt and Robert (1990, 1994)]. Given $\mathbf{y} = (y_1, \ldots, y_n)$ from (2.1), we can associate to each $y_i$ an integer $x_i$ between 1 and $J$ that identifies the component of the mixture generating observation $y_i$. Thus, the variable $x_i$ takes value $j$ with prior probability $\pi_j$, $1 \le j \le J$. The vector $\mathbf{x} = (x_1, \ldots, x_n)$ of component labels is the *missing data* part of the sample since it is not observed. Figure 1 illustrates this for our model, highlighting that $\mathbf{y}$ is conditionally independent from the mixture weights $\boldsymbol{\pi}$, given the missing data $\mathbf{x}$.

Suppose the missing data $x_1, \ldots, x_n$ were available. Then the model could be written as

$$(2.3) \quad p(y_1, \ldots, y_n | x_1, \ldots, x_n, \theta) = \frac{\theta^{\sum_{i=1}^{n} x_i}}{\prod_{i=1}^{n} \Gamma(x_i)} \left(\prod_{i=1}^{n} y_i^{x_i - 1}\right) e^{-\theta \sum_{i=1}^{n} y_i}.$$

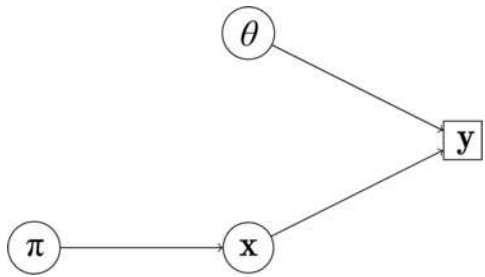

FIG. 1. *Directed acyclic graph (DAG) for the missing data representation of the $\mathcal{GSM}(\boldsymbol{\pi}, \theta | J)$ model.*

Thus, using (2.3) and the priors, the posterior distribution is

$$(2.4) \quad \begin{aligned} &p(\pi_1, \ldots, \pi_J, \theta | y_1, \ldots, y_n, x_1, \ldots, x_n) \\ &\propto \left( \prod_{j=1}^{J} \pi_j^{(1/J) + n_j - 1} \right) \theta^{\alpha + (\sum_{i=1}^{n} x_i) - 1} e^{-(\beta + \sum_{i=1}^{n} y_i)\theta}, \end{aligned}$$

where $n_j = \sum_{i=1}^{n} \mathbb{I}(x_i = j)$, $j = 1, \ldots, J$, and $\mathbb{I}(\cdot)$ is the indicator function. The main consequence of this conditional decomposition is that, for a given missing data vector $x_1, \ldots, x_n$, the conjugacy is preserved and, therefore, the simulation can be easily performed, conditional on the missing data $x_1, \ldots, x_n$.

2.2. *Computation of the posterior distribution.* We implement two approaches for estimating the unknown parameters of interest. In the first approach we estimate the posterior distribution of $\boldsymbol{\pi}$, $\mathbf{x}$ and $\theta$ by using a Gibbs sampler (details are reported in the Appendix). To increase the efficiency, we also propose a second estimation approach where we integrate out the scale parameter $\theta$ analytically. The advantage of this second strategy is both computational, since the Markov chain runs in a smaller space, and theoretical, since generally simulated values are less autocorrelated after partial marginalization [Liu (1994), MacEachern (1994), MacEachern, Clyde and Liu (1999)].

After having integrated out $\theta$, the full conditional distribution of the mixture weights is given by

$$p(\pi_1, \ldots, \pi_J | y_1, \ldots, y_n, x_1, \ldots, x_n) \propto \prod_{j=1}^{J} \pi_j^{(1/J) + n_j - 1},$$

that is, the $\mathcal{D}_J(\frac{1}{J} + n_1, \ldots, \frac{1}{J} + n_J)$ Dirichlet distribution. In addition, the full conditional of the $i$th missing label is then given by

$$(2.5) \quad p(x_i | \mathbf{y}, \mathbf{x}_{(-i)}, \boldsymbol{\pi}) = \sum_{j=1}^{J} \frac{\pi_j \frac{y_i^{j-1}}{\Gamma(j)} \frac{(\alpha + \sum_{(-i)} x_r)_j}{(\beta + \sum_{r=1}^{n} y_r)^j}}{\sum_{k=1}^{J} \pi_k \frac{y_i^{k-1}}{\Gamma(k)} \frac{(\alpha + \sum_{(-i)} x_r)_k}{(\beta + \sum_{r=1}^{n} y_r)^k}} \mathbb{I}(x_i = j),$$

where $\mathbf{x}_{(-i)}$ is the $\mathbf{x} = (x_1, \ldots, x_n)$ vector with the $i$th element deleted, $\sum_{(-i)} x_r$ denotes the sum of all the component labels except for the $i$th one, and $(n)_k = n(n+1)\cdots(n+k-1)$ is the Pochhammer symbol [see Abramowitz and Stegun (1972)]. We assume that $\alpha$ is an integer, for computation speed, and to avoid overflow errors. See Section A.1 in the Appendix for further details. The integration of $\theta$ implies that the missing data are no longer independent.

We have implemented this approach in an R package called GSM [Venturini, Dominici and Parmigiani (2008)], also available from the Comprehensive R Archive Network (www.r-project.org).

2.3. *Choice of hyperparameters.* Our model specification only requires the user to specify three hyperparameters: the number of components $J$, and the $\alpha$ and $\beta$ from the conjugate prior on $\theta$. Though these can be chosen based on prior knowledge when available, we also propose, and use in our application, an informal Empirical Bayes approach which uses summary statistics of the data, like the maximum and the sum of the observations, to get reasonable values.

The mean of a GSM random variable with distribution (2.1) is

$$(2.6) \qquad \mu = \mathbb{E}[Y|\pi_1, \ldots, \pi_J, \theta] = \sum_{j=1}^{J} \pi_j \frac{j}{\theta},$$

so the scale parameter $\theta$ can be expressed as

$$(2.7) \qquad \theta = \frac{1}{\mu} \sum_{j=1}^{J} \pi_j j,$$

that is, a weighted average of the $J$ shape parameters divided by the mean. Also, the expected value of the full conditional distribution of $\theta$ is

$$
\begin{aligned}
(2.8) \qquad \mathbb{E}[\theta|\mathbf{y}, \mathbf{x}] &= \frac{\alpha + \sum_{i=1}^{n} x_i}{\beta + \sum_{i=1}^{n} y_i} \\
&= \frac{\beta}{\beta + \sum_{i=1}^{n} y_i} \cdot \frac{\alpha}{\beta} + \frac{\sum_{i=1}^{n} y_i}{\beta + \sum_{i=1}^{n} y_i} \cdot \frac{\sum_{i=1}^{n} x_i}{\sum_{i=1}^{n} y_i} \\
&= \omega \cdot \frac{\alpha}{\beta} + (1 - \omega) \cdot \frac{\bar{x}}{\bar{y}}.
\end{aligned}
$$

Equation (2.8) indicates that the posterior mean of $\theta$ is a weighted average of $\alpha/\beta$, the prior mean of $\theta$, and $\bar{x}/\bar{y}$. When both $\alpha \to 0$ and $\beta \to 0$ the prior becomes improper. Then, for a given value of $J$, a strategy for choosing $\alpha$ and $\beta$ is as follows:

1. Compute $\widetilde{\theta} = J/\max(y_1, \ldots, y_n)$ and check that $1/\widetilde{\theta} \leq \min(y_1, \ldots, y_n)$; the idea is that, on average, $\theta$ should take values that allow the set of gamma distributions in (2.1) to completely span the range of observed values (the last gamma distribution should have a mean not smaller than the maximum observation and the first gamma distribution a mean not greater than the minimum observation). $\widetilde{\theta}$ is hence a candidate for the prior mean $\alpha/\beta$.

2. Choose a value for the weight of the prior information $\omega$ in (2.8). Values between 0.2 and 0.5 are usually reasonable choices. Fix $\beta$ to $(\omega \cdot \sum_{i=1}^{n} y_i)/(1 - \omega)$.

3. Set $\alpha$ by rounding to the closest integer the quantity $\widetilde{\theta} \cdot \beta$. The rounding is needed because of the assumption used to get (2.5).

Concerning the choice of $J$, the goodness of fit of the GSM is the result of the interplay among the grids of $m$th order moments

$$\left(\frac{\prod_{\ell=1}^{m} \ell}{\theta^m}, \frac{\prod_{\ell=1}^{m}(\ell+1)}{\theta^m}, \ldots, \frac{\prod_{\ell=1}^{m}(\ell+J-2)}{\theta^m}, \frac{\prod_{\ell=1}^{m}(\ell+J-1)}{\theta^m}\right),$$

and the ordered sequence of observations. These grids should contain sufficient elements to fit the data, therefore, $J$ should be calibrated to the specific set of data under examination. Generally speaking, a small value of $J$ can create a severe limitation to the model, as the set of densities available in the class being mixed may not be sufficiently rich with elements that have a large mean. On the other hand, a too large value does not cause serious difficulties as the fit is often robust when there are several gamma distributions in the class that can serve as building blocks for a particular mixture component. However, large $J$ can cause numerical problems. Sometimes a transformation of the data (like a log or a root) can be useful to handle these numerical issues. In practice, the choice of $J$ may require more than one iteration. Inspection of the predictive density is a practical diagnostic to identify misspecification of $J$.

**3. Results.** In this section we carry out a simulation study to assess the predictive performance of the gamma mixture model in a controlled situation in which a large test set is available to serve as the gold standard. Our simulation reflects closely the estimation of the right tail of the medical expenditures distribution for smoking attributable diseases (CHD and LC) from the MCBS. We compare our model with other popular methods in health services research. The results show that these standard approaches are less effective in real life situations. In addition, we apply our methods to the MCBS data for estimating the risk for persons affected by smoking attributable diseases to exceed a given medical costs threshold in a single hospitalization.

3.1. *Simulation design.* From the complete MCBS data for the period 1999–2002, we extract expenditure data on hospitalizations in which the first diagnosis has been CHD, LC, or both, for a total of 7,615 hospitalizations. Tables 1 and 2 report a brief summary of the dataset. From this population, we drew 500 sub-samples, the *training sets*, of size equal to 10% of the original sample. For each draw, the remaining 90% constitute the *test set*.

On each training set we estimate the tail probability $\hat{p} = \mathbb{P}(y^* > k|\mathbf{y})$ using the following four approaches:

EDF  The empirical proportion of cases exceeding the threshold, or the empirical distribution function (EDF), which provides a straightforward model-free estimate;

LN The area to the right of $k$ in a log-normal distribution (LN) whose parameters are estimated by maximum likelihood [see Aitchison and Shen (1980), Zellner (1971)]; this method is commonly used in health economics and provides a benchmark to assess the gains provided by our approach compared to standard practice;

MN The area to the right of $k$ in a mixture of normal distributions (MN), estimated assuming an unknown number of components using the `mclust` package in `R` [see Fraley and Raftery (2002, 2006)]; this is a flexible and commonly used mixture model and provides a benchmark to assess the gains provided by our novel mixture approach compared to existing mixture approaches;

GSM The gamma shape mixture distribution (GSM) estimator proposed here.

For the GSM, the estimator of the tail probability is defined as

$$(3.1) \qquad \mathbb{P}(y^* > k|\mathbf{y}) = \int \mathbb{P}(y^* > k|\mathbf{y}, \theta, \boldsymbol{\pi}) f(\theta, \boldsymbol{\pi}|\mathbf{y}) \, d\theta \, d\boldsymbol{\pi}.$$

This predictive probability can be estimated from Gibbs sampling realizations by the Rao–Blackwellized estimator:

$$\widehat{\mathbb{P}}(y^* > k|\mathbf{y}) = \frac{1}{M} \sum_{m=1}^{M} \mathbb{P}(y^* > k|\theta^{(m)}, \boldsymbol{\pi}^{(m)})$$

$$= \frac{1}{M} \sum_{m=1}^{M} \sum_{j=1}^{J} \pi_j^{(m)} [1 - F_j(k|\theta^{(m)})],$$

where $F_j(\cdot|\theta)$ is the distribution function of a $\mathcal{G}a(j, \theta)$ random variable. In this analysis the hyperparameters for the GSM are $J = 200$, $\alpha = 12{,}380$, $\beta = 3{,}420$, and have been chosen following the indications provided in Section 2.3. We run the chain for 5,000 iterations (1,000 of which for burn-in).

On each test set we compute the proportion $p_{\text{TRUE}}$ of expenditures exceeding $k$, and use it as the gold standard. We repeat the entire analysis for the following values of the medical cost threshold $k$: \$10,000, \$15,000, \$20,000, \$30,000, \$50,000 and \$80,000 (higher threshold values are too rare to be worth including in the analysis, see Table 2).

TABLE 1
*Summary of the MCBS dataset: high-order quantiles of medical expenditures of first hospitalization for LC, CHD or both*

| Quantile order | 75 | 90 | 95 | 97.5 | 99 | 99.9 |
|---|---|---|---|---|---|---|
| Quantile (\$) | 8,187.5 | 15,457.2 | 22,485.6 | 29,009.4 | 40,955.9 | 115,060.6 |

TABLE 2
*Summary of the MCBS dataset: number of hospitalizations with a cost above a specified threshold*

| Threshold ($) | 10,000 | 15,000 | 20,000 | 30,000 | 50,000 | 80,000 | 100,000 |
|---|---|---|---|---|---|---|---|
| Count | 1,468 | 799 | 503 | 179 | 48 | 24 | 9 |
| Proportion | 0.1928 | 0.1049 | 0.0661 | 0.0235 | 0.0063 | 0.0032 | 0.0012 |

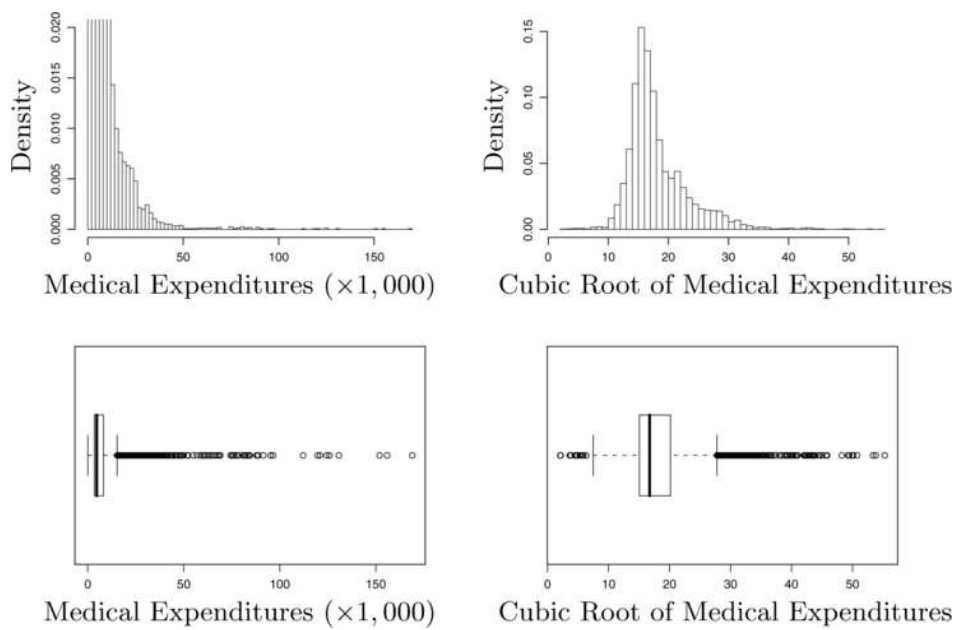

FIG. 2. *Histograms and boxplots of positive Medicare expenditures for hospitalizations regarding smoking attributable diseases (lung cancer and coronary heart disease) from the 1999–2002 Medicare Current Beneficiaries Survey (for clarity of exposition, the histogram of the original expenditures has been truncated at the top).*

The data are transformed using the cubic root to control numerical overflow. The population parameter of interest remains unchanged by this transformation. Figure 2 reports graphical summaries for both the untransformed and transformed MCBS data.

3.2. *Simulation results.* Figure 3 summarizes the results of our simulation, for each estimation method and each $k$. In panel (a) we present the relative mean squared errors, defined by $(\mathrm{mse}_{\mathrm{EDF}} - \mathrm{mse}_{\hat{p}})/\mathrm{mse}_{\mathrm{EDF}}$, where $\mathrm{mse}_{\hat{p}}$ indicates the mean squared error for the tail probability $\mathbb{P}(y^* > k|\mathbf{y})$ estimated using tail probability estimator $\hat{p}$—either LN, MN or GSM. The relative mean square errors are obtained as the ratios of the averaged mean squared errors over the 500 sub-samples. In panel (b) we present the relative bias $(\mathbb{E}(\hat{p}) - p_{\text{TRUE}})/p_{\text{TRUE}}$. Both ratios are multiplied by 100. Negative values of the relative mean squared error imply that the EDF estimator is preferred to $\hat{p}$, while positive values imply that the tail probability estimator $\hat{p}$ is preferred to the EDF.

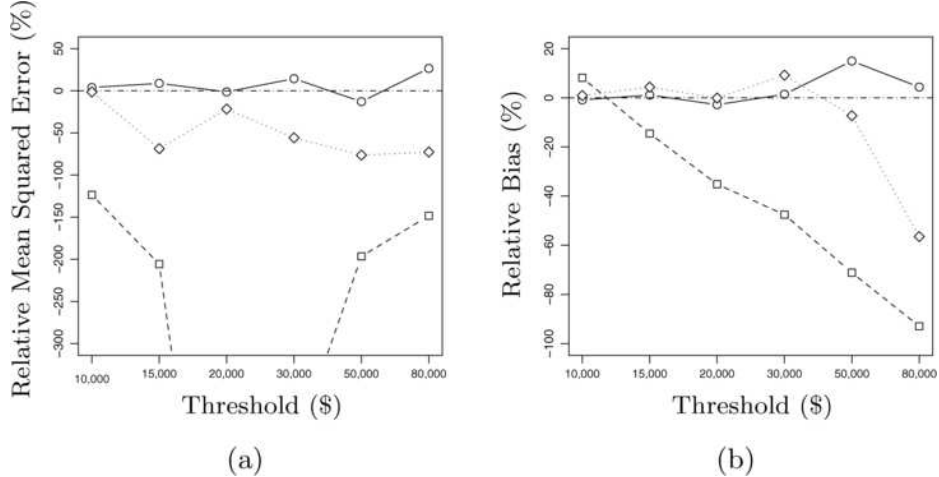

(a)                                    (b)

FIG. 3. *Simulation study.* (a) *For different threshold values, the percent mean squared error relative to the EDF estimator, that is,* $[(\text{mse}_{\text{EDF}} - \text{mse}_{\hat{p}})/\text{mse}_{\text{EDF}}] \times 100$. (b) *For different threshold values, the percent bias relative* $p_{\text{TRUE}}$, *that is,* $[(\mathbb{E}(\hat{p}) - p_{\text{TRUE}})/p_{\text{TRUE}}] \times 100$. *In both panels, circles indicate GSM, diamonds MN and squares LN.*

For almost all medical expenditure thresholds, the GSM is more efficient than the estimator based on the empirical distribution function. As expected, we observe a trend for the relative efficiency to increase with the threshold, up to a 27% improvement for the highest threshold.

The tail probability estimator based on the log-normal distribution performs poorly. Its mean squared error relative to the EDF estimator is below $-100\%$ for all the thresholds, with the worst performance corresponding to hospitalization costs above \$20,000. This result is important practically since many models in health services research and health economics are based on the assumption that the medical expenditures can be handled using log-normal distributions.

Also, the estimator based on the mixture of normal distributions performs worse than the EDF, and the performance worsens as the threshold increases. This finding suggests that normal mixtures, while flexible, are not completely adequate for heavily skewed data when the focus of the investigation is a small tail probability.

Figure 3(b) reports the relative biases, showing that for large $k$ the mixture of gamma distributions is slightly biased, but less so than the other methods. Once again, the estimator based on the log-normal distribution is more biased than the alternatives, and it almost always underestimates the tail probability for the reference population. The bias increases systematically (in absolute value) with the thresholds, indicating that the log-normal distribution is not sufficiently heavy-tailed to mimic the right tail of these data.

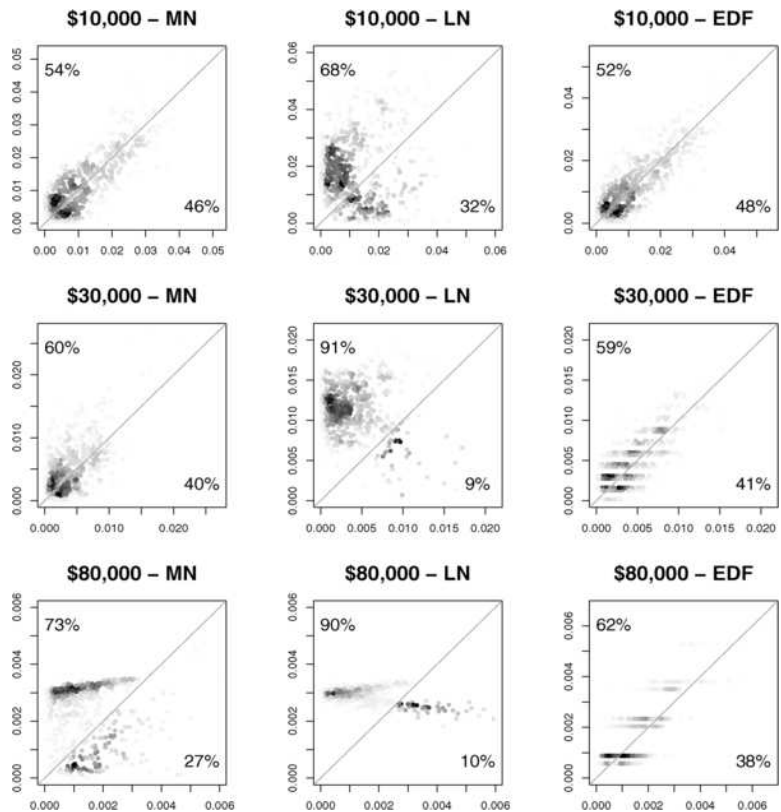

Fig. 4. *Simulation study: pairwise comparison of predictive performances. Each point represents a sub-sample of the simulation study. The coordinates of each point are given by the absolute value of the difference between the estimated tail probability on the training and the gold standard, for the GSM estimator (on the horizontal axes) and for an alternative estimator (on the vertical axes), as indicated on top of each plot. Graphs on different rows refer to different medical expenditure thresholds. The shading shows the density of the points. Also indicated are the percentages of points above and below the 45 degrees line.*

Figure 4 allows a further comparison of the GSM with the other models. In each panel we show scatterplots of the performance of GSM versus that of the other estimation methods for three thresholds: $10,000, $30,000 and $80,000. A point in these graphs represents a single sub-sample of the simulation study. Within each panel, the x-axis is the absolute value of the difference

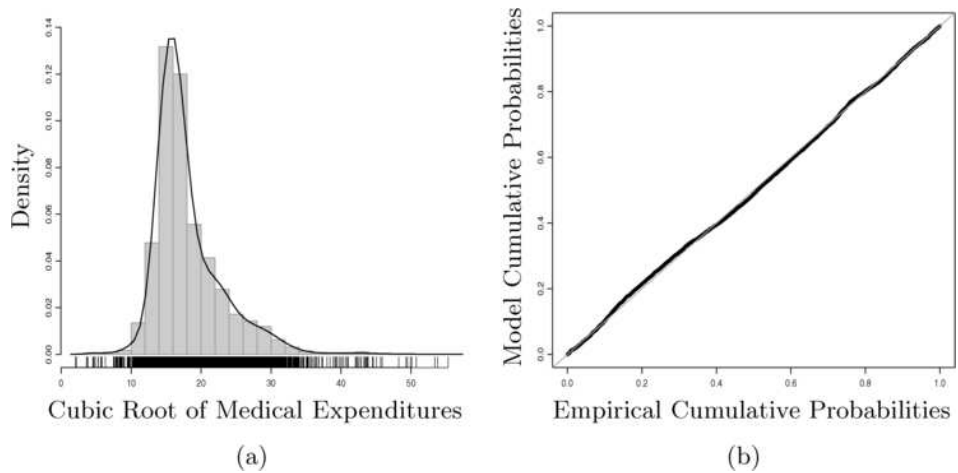

Fɪɢ. 5.  *MCBS data analysis. Fit of the GSM model to the MCBS medical costs related to smoking attributable diseases ($n = 7,615$ hospitalizations). (a) Histogram of the cubic root transformed medical costs together with the posterior mean of the model density (solid line). (b) QQ-plot of the estimated model cumulative probabilities versus the empirical ones.*

between the estimated tail probability on the training set using the GSM and the gold standard; the y-axis is computed in the same fashion, but for the alternative method indicated on top of the panel. These graphs suggest that the GSM performs better than the other estimation methods from a predictive point of view, since in every panel the majority of points are above the diagonal. The higher the threshold, the more pronounced the result. Moreover, even if slightly biased, most of the times[1] the GSM performs better than the estimator based on the empirical distribution function.

3.3. *Analysis of MCBS medical costs data.*  In this section we present a data analysis of the MCBS dataset. The goals of the analysis are to provide estimates of the density function and of the risk of exceeding a given medical cost threshold $k$ in a single hospitalization, with associated probability intervals. As in the simulation study, we restrict the analysis to hospitalizations in which the first diagnosis has been for a smoking attributable disease, CHD or LC. The size of the sample is $n = 7,615$.

The hyperparameters for the GSM have been set to $J = 200$, $\alpha = 124{,}960$, $\beta = 34{,}520$. The data have been transformed using a cubic root transformation. We used 6,000 sampling iterations (1,000 of which as burn-in).

---

[1]In Figure 4 only the graphs for some of the available thresholds are shown to avoid cluttering.

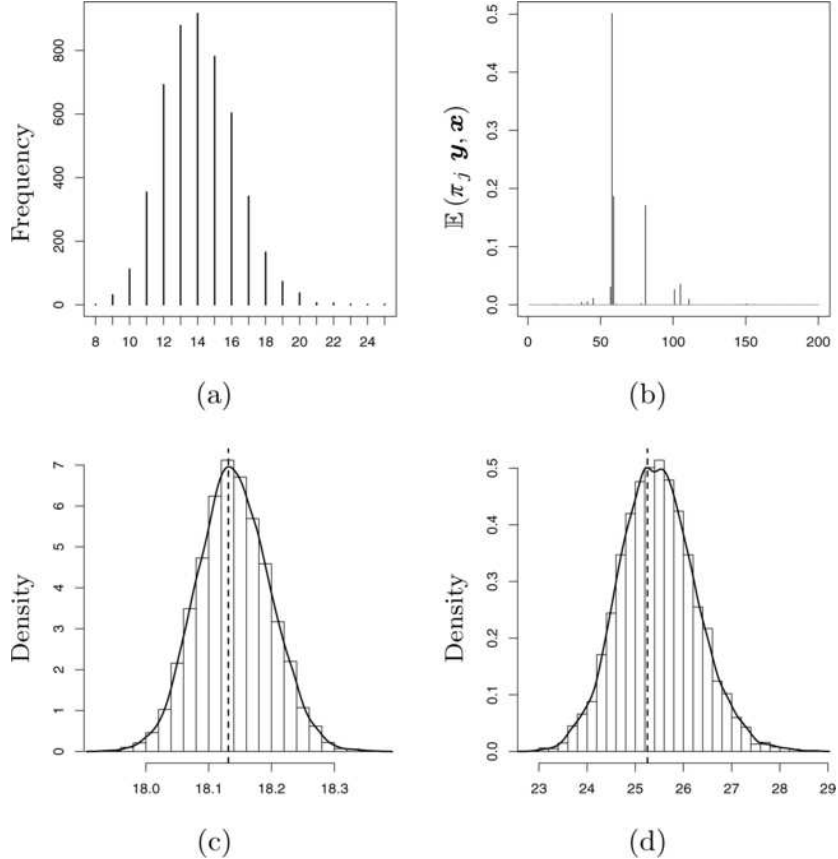

Fig. 6. *MCBS data analysis.* (a) *Distribution of the number of mixture components represented in the sample at each iteration of the chain;* (b) *Posterior mean of the mixture weights;* (c) *Posterior distribution of the model mean [see (2.6)]. The vertical dashed line is the sample mean and the solid line is the kernel density estimator;* (d) *Posterior distribution of the model variance. The vertical dashed line indicates the data sample variance, while the over-imposed solid line is the kernel density estimator. The last two panels are useful checks of whether the key quantities $\alpha$, $\beta$ and $J$ have been chosen appropriately.*

Figure 5 shows the fit of the GSM model to the MCBS data. Panel (a) presents the data histogram together with the fitted model density, estimated as the average over the posterior distribution of the model parameters $\boldsymbol{\pi}$ and $\theta$ of the mixture density evaluated at a grid of points. Panel (b) reports the QQ-plot of the model cumulative probabilities, evaluated at the posterior mean of the mixture weights and scale parameter, versus the empirical cumulative probabilities $p_i = i/(n+1)$, $i = 1, \ldots, n$. As it is clear from these graphs, the GSM provides a very good representation of the data.

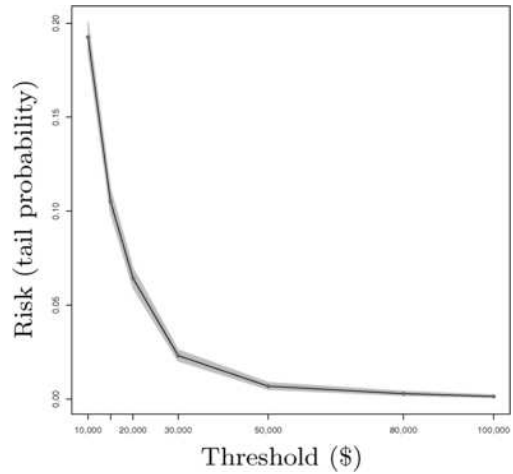

Fig. 7. *MCBS data analysis. Risk to exceed a given medical costs threshold in a single hospitalization. Each point corresponds to the estimate of the predictive posterior probability $\hat{\mathbb{P}}(y^* > k|\mathbf{y})$ obtained with the GSM model on the MCBS data. Shading represents the 95% credible intervals.*

Figure 6 contains additional results that provide important insight on how the model works. Even though $J = 200$ components were available, the estimation procedure selects, at every iteration, only a small subset, sufficient to fit the data. Panel (a) shows that, a posteriori, the number of selected components is between 9 and 20, with mode at 14. Panel (b) shows the posterior means of the mixture weights: only few components have a posterior mean weight that is substantially greater than zero. This suggests that despite the availability of $J$ parameters, the posterior concentrates on a much smaller subspace. If desired, a simple hard thresholding of the posterior probabilities could produce a parsimonious representation of the model based on few mixture components. Panels (c) and (d) report the histograms of the model mean and variance evaluated at each Gibbs sampler iteration. The vertical dashed lines indicate the empirical sample mean and variance. These plots are useful to qualitatively assess whether hyperparameters $\alpha$, $\beta$ and $J$ are appropriate for the sample at hand.

Figure 7 displays the estimates of the tail probability for different threshold values. In other words, this graph presents the "risk" of exceeding a given medical costs threshold in a single hospitalization for people affected by smoking attributable diseases. The 95% credible intervals for each threshold estimate are also shown.

We also performed a set of sensitivity analyses to prior hyperparameters, by varying $\omega = \beta/(\beta + \sum_{i=1}^{n} y_i)$ in the range 0.2–0.5 and found that the results are not significantly affected.

**4. Discussion.** In this paper we introduced a Bayesian mixture model for density and tail probability estimation of very skewed distributions. Our approach is based on a mixture of gamma distributions over the shape parameter, or GSM. This family of distributions includes components whose means and variances increase together, offering a convenient way of representing populations in which a small fraction of individuals has an outlying behavior that is difficult to predict.

We also develop a computationally efficient algorithm which is a modification of the standard Gibbs sampler used in the Bayesian literature on mixture distributions [Robert (1996)]. In our approach we integrate out the scale parameter $\theta$, thus making computations more efficient [Liu (1994)]). Convergence of our algorithm is, in our experience so far, fast and reliable. No tuning parameters are required. Our implementation is available as an R package called GSM, which can be obtained from the Comprehensive R Archive Network (www.r-project.org).

Our approach was motivated by the estimation of the proportion of subjects affected by smoking attributable diseases, specifically CHD and LC, that, in a single hospitalization, have a medical bill exceeding a given threshold. In particular, we used data from MCBS, a multipurpose survey of a nationally representative sample of Medicare beneficiaries in the U.S. However, in different contexts, one may consider taking into account the actual survey weights to get more representative estimates of the medical expenditures extreme percentiles. The appropriate utilization of such adjustment is still debated in the literature [see, e.g., Gelman (2007), Lohr (1999) and Pfeffermann (1993)].

The large survey considered in our article allowed us to perform a rigorous cross-validation experiment in which very large test sets provide a reliable gold standard. In this context we demonstrated that GSM offer a far more accurate prediction than that provided by the log-normal model, commonly utilized in health economics and health services research. We conclude that for highly skewed data the log-normal model is unlikely to be appropriate and unchecked use can lead to seriously biased and inefficient estimates of tail probability, especially for high thresholds. The simulation shows that GSM is also competitive compared to flexible alternatives such as normal mixture models and model-free estimators. Additionally, being a complete probabilistic specification of the data generation process, it can be used for any other inferential purpose.

Fitting a GSM model requires selecting the maximum number of mixture components $J$. A possible generalization is to incorporate the model into a reversible jump approach, as proposed by Richardson and Green (1997), in which the number of components is random. This would come at the cost of a substantial increase in the computing time, length of the required chains, and convergence monitoring efforts. However, because of the limited loss in specifying a large $J$, the tool we described is likely to be very useful as currently proposed. In our model, $J$ does not have to fit the data: it only has to be large enough to contain all the necessary components to fit the data. In this sense, choosing $J$ is more similar to choosing a model space than a model. Also, not having to implement and monitor an RJ-MCMC makes our approach more suitable for off-the-shelf use of the associated software.

Our presentation considered a relatively simple context in which all subjects belong to the same group and are measured once. The logic of our approach lends itself to generalizations to more complex scenarios. First, it may be useful to model the scale parameter as a function of covariates of interest. This extension would allow to get a general and robust regression model for skewed data. Second, in medical cost data, there may be availability of multiple hospitalizations records for some subjects in the survey. Including this dependence in the model would allow for more precise individual-level predictions. Finally, one may need to incorporate censoring. In cost data, censoring can be present because of caps in reimbursement [see, e.g., Lipscomb et al. (1998)]. Also, if cost was not defined for a single hospitalization but referred to an extended period, traditional time-to-event censoring would come into play as well.

## APPENDIX: COMPUTATIONAL DETAILS

**A.1. Gibbs sampler for mixture estimation with $\theta$ integrated out.** The posterior distribution of $(\pi_1, \ldots, \pi_J, \theta)$, given the sample $(y_1, \ldots, y_n)$, can be written as

$$p(\pi_1, \ldots, \pi_J, \theta | y_1, \ldots, y_n)$$
$$\propto \left( \prod_{j=1}^{J} \pi_j^{(1/J)-1} \right) \theta^{\alpha-1} e^{-\beta\theta} \prod_{i=1}^{n} \left( \sum_{j=1}^{J} \pi_j \frac{\theta^j}{\Gamma(j)} y_i^{j-1} e^{-\theta y_i} \right).$$

The standard algorithm to implement the posterior simulation is reported in the next subsection. However, to increase efficiency, in our estimation approach we integrate out the scale parameter $\theta$ [Liu (1994), MacEachern (1994), MacEachern, Clyde and Liu (1999)]. Then (2.3) becomes

$$p(y_1, \ldots, y_n | x_1, \ldots, x_n)$$

(A.1)
$$= \int_0^\infty \frac{\theta^{\sum_{i=1}^n x_i}}{\prod_{i=1}^n \Gamma(x_i)} \left( \prod_{i=1}^n y_i^{x_i-1} \right) e^{-\theta \sum_{i=1}^n y_i} \frac{\beta^\alpha}{\Gamma(\alpha)} \theta^{\alpha-1} e^{-\beta\theta} \, d\theta$$

$$= \frac{\beta^\alpha}{\Gamma(\alpha)} \frac{\prod_{i=1}^n y_i^{x_i-1}}{\prod_{i=1}^n \Gamma(x_i)} \int_0^\infty \theta^{\alpha+(\sum_{i=1}^n x_i)-1} e^{-(\beta+\sum_{i=1}^n y_i)\theta} \, d\theta$$

$$= \frac{\beta^\alpha}{\Gamma(\alpha)} \frac{\prod_{i=1}^n y_i^{x_i-1}}{\prod_{i=1}^n \Gamma(x_i)} \frac{\Gamma(\alpha+\sum_{i=1}^n x_i)}{(\beta+\sum_{i=1}^n y_i)^{\alpha+(\sum_{i=1}^n x_i)}}.$$

Note that the observed data, conditionally on the nonobserved ones, are no longer independent. The interpretation of this fact is that $\theta$ was a parameter shared by all the $(y_i, x_i)$ pairs, $i = 1, \ldots, n$. Removing $\theta$ has introduced dependence among the data.

The full conditional of the mixture weights is hence given by

$$p(\pi_1, \ldots, \pi_J | y_1, \ldots, y_n, x_1, \ldots, x_n) \propto \prod_{j=1}^{J} \pi_j^{(1/J) + n_j - 1},$$

while, to get the full conditional of the missing data, we decompose it into the individual full conditionals

$$p(x_i | y_1, \ldots, y_n, x_1, \ldots, x_{i-1}, x_{i+1}, \ldots, x_n, \pi_1, \ldots, \pi_J),$$

$i \in \{1, \ldots, n\}$. Note that

$$p(x_i | \mathbf{y}, \mathbf{x}_{(-i)}, \boldsymbol{\pi}) = \sum_{j=1}^{J} \frac{p(x_i, \mathbf{x}_{(-i)} | \mathbf{y}, \boldsymbol{\pi})}{\sum_{k=1}^{J} p(k, \boldsymbol{x}_{(-i)} | \mathbf{y}, \boldsymbol{\pi})} \mathbb{I}(x_i = j)$$

$$= \sum_{j=1}^{J} \frac{p(\mathbf{y} | x_i, \mathbf{x}_{(-i)}) \cdot p(x_i, \mathbf{x}_{(-i)} | \boldsymbol{\pi})}{\sum_{k=1}^{J} p(\mathbf{y} | k, \mathbf{x}_{(-i)}) \cdot p(k, \mathbf{x}_{(-i)} | \boldsymbol{\pi})} \mathbb{I}(x_i = j),$$

for $i \in \{1, \ldots, n\}$ and where $\mathbf{x}_{(-i)}$ denotes the $\mathbf{x} = (x_1, \ldots, x_n)$ vector with the $i$th element deleted. The second equality follows from $p(\mathbf{x} | \mathbf{y}, \boldsymbol{\pi}) \cdot p(\mathbf{y} | \boldsymbol{\pi}) = p(\mathbf{y} | \mathbf{x}, \boldsymbol{\pi}) \cdot p(\mathbf{x} | \boldsymbol{\pi})$ and from the conditional independence of $\mathbf{y}$ from $\boldsymbol{\pi}$, given the missing data $\mathbf{x}$. Substituting (A.1), we obtain

$$(A.2) \quad p(x_i | \mathbf{y}, \mathbf{x}_{(-i)}, \boldsymbol{\pi}) = \sum_{j=1}^{J} \frac{\pi_j \frac{y_i^{j-1}}{\Gamma(j)} \frac{\Gamma(\alpha + \sum_{(-i)} x_r + j)}{(\beta + \sum_{r=1}^{n} y_r)^j}}{\sum_{k=1}^{J} \pi_k \frac{y_i^{k-1}}{\Gamma(k)} \frac{\Gamma(\alpha + \sum_{(-i)} x_r + k)}{(\beta + \sum_{r=1}^{n} y_r)^k}} \mathbb{I}(x_i = j),$$

where the $\sum_{(-i)} x_r$ denotes the sum of all the component labels apart from the $i$th one. If one further assumes that $\alpha \in \mathbb{N}$, then (A.2) can be further simplified to

$$(A.3) \quad p(x_i | \mathbf{y}, \mathbf{x}_{(-i)}, \boldsymbol{\pi}) = \sum_{j=1}^{J} \frac{\pi_j \frac{y_i^{j-1}}{\Gamma(j)} \frac{(\alpha + \sum_{(-i)} x_r)_j}{(\beta + \sum_{r=1}^{n} y_r)^j}}{\sum_{k=1}^{J} \pi_k \frac{y_i^{k-1}}{\Gamma(k)} \frac{(\alpha + \sum_{(-i)} x_r)_k}{(\beta + \sum_{r=1}^{n} y_r)^k}} \mathbb{I}(x_i = j),$$

where $(n)_k = n(n+1) \cdots (n+k-1)$ is the Pochhammer symbol [see Abramowitz and Stegun (1972)]. We refer to the whole fraction inside the leftmost summation as $\kappa_{ij}$. In principle, $\alpha$ could be any positive real number, but constraining it to be an integer allows us to simplify large quantities from the numerator and denominator of (A.2) and, hence, to easily prevent overflow errors during the calculations.

The steps to implement this simulation algorithm are then summarized below:

- Simulate

$$\boldsymbol{\pi}|\mathbf{y},\mathbf{x} \sim \mathcal{D}_J\left(\frac{1}{J}+n_1,\dots,\frac{1}{J}+n_J\right),$$

  where $n_j = \sum_{i=1}^n \mathbb{I}(x_i = j)$, $j = 1,\dots,J$.
- Simulate, for every $i = 1,\dots,n$,

$$p(x_i|y_1,\dots,y_n,x_1,\dots,x_{i-1},x_{i+1},\dots,x_n,\pi_1,\dots,\pi_J) = \sum_{j=1}^J \kappa_{ij}\mathbb{I}(x_i = j),$$

  with $\kappa_{ij}$ as defined above, $j = 1,\dots,J$.
- Update $n_j$, $j = 1,\dots,J$.

**A.2. Standard Gibbs sampler for mixture estimation.** The implementation of the standard Gibbs sampling is straightforward and involves the iterative simulation from (2.4), for the parameters of the model, and from

$$p(x_1,\dots,x_n|\pi_1,\dots,\pi_J,\theta,y_1,\dots,y_n),$$

for the missing data. The steps for the algorithm are [see, e.g., Robert (1996)]:

- Simulate

$$\theta|\mathbf{y},\mathbf{x},\boldsymbol{\pi} \sim \mathcal{G}a\left(\alpha + \sum_{i=1}^n x_i, \beta + \sum_{i=1}^n y_i\right),$$

$$\boldsymbol{\pi}|\mathbf{y},\mathbf{x},\theta \sim \mathcal{D}_J\left(\frac{1}{J}+n_1,\dots,\frac{1}{J}+n_J\right),$$

  where $n_j = \sum_{i=1}^n \mathbb{I}(x_i = j)$, $j = 1,\dots,J$.
- Simulate, for every $i = 1,\dots,n$,

$$p(x_i|y_i,\pi_1,\dots,\pi_J,\theta) = \sum_{j=1}^J \pi_{ij}\mathbb{I}(x_i = j),$$

  where

$$\pi_{ij} = \frac{\pi_j f_j(y_i|\theta)}{\sum_{k=1}^J \pi_k f_k(y_i|\theta)}, \qquad j = 1,\dots,J.$$

- Update $n_j$, $j = 1,\dots,J$.

**Acknowledgments.** We would like to thank the reviewers for their helpful comments.

## SUPPLEMENTARY MATERIAL

**Gamma shape mixture** (doi: [10.1214/08-AOAS156SUPP](10.1214/08-AOAS156SUPP)). This package implements a Bayesian approach for estimation of a mixture of gamma distributions in which the mixing occurs over the shape parameter. This family provides a flexible and novel approach for modeling heavy-tailed distributions, it is computationally efficient, and it only requires to specify a prior distribution for a single parameter.

S. VENTURINI
DIPARTIMENTO DI SCIENZE DELLE DECISIONI
UNIVERSITÀ COMMERCIALE LUIGI BOCCONI
VIALE ISONZO 25
MILANO 20137
ITALY
E-MAIL: sergio.venturini@unibocconi.it

F. DOMINICI
DEPARTMENT OF BIOSTATISTICS
JOHNS HOPKINS SCHOOL OF PUBLIC HEALTH
615 NORTH WOLFE STREET
BALTIMORE, MARYLAND 21205
USA

G. PARMIGIANI
THE SIDNEY KIMMEL COMPREHENSIVE CANCER CENTER
JOHNS HOPKINS UNIVERSITY
550 NORTH BROADWAY
BALTIMORE, MARYLAND 21205
USA